\documentclass[12pt]{article}
\usepackage{amsmath}
\usepackage{amssymb}
\usepackage{cancel}
\usepackage{graphicx}
\usepackage{wrapfig}

\numberwithin{equation}{section}

\begin{document}
\baselineskip=19pt

\begin{titlepage}

\begin{center}
\vspace*{17mm}

{\large\bf%
Neutrino masses and mixing from $S_4$ flavor twisting
}

\vspace*{10mm}
Hajime~Ishimori,$^{*}$
~Yusuke~Shimizu,$^{*}$\\
Morimitsu~Tanimoto,$^{\dag}$
~Atsushi Watanabe$^\dag$ 
\vspace*{10mm}

$^\dag${\small {\it Graduate~School~of~Science~and~Technology,~Niigata~University, \\ 
Niigata~950-2181,~Japan, }} \\
$^\dag${\small {\it Department of Physics, 
Niigata University, Niigata 950-2181, Japan}}\\

\vspace*{3mm}

{\small (October, 2010)}
\end{center}

\vspace*{7mm}

\begin{abstract}\noindent%
We discuss a neutrino mass model based on the $S_4$ discrete symmetry
where the symmetry breaking is triggered by the boundary conditions of 
the bulk right-handed neutrino in the fifth spacial dimension.
The three generations of the left-handed lepton doublets and the 
right-handed neutrinos are assigned to be the triplets of $S_4$.
The magnitudes of the lepton mixing angles, especially the reactor angle is 
related to the neutrino mass patterns and the model will be tested in future 
neutrino experiments, e.g., an early discovery of the reactor angle favors 
the normal hierarchy. For the inverted hierarchy, the lepton mixing is 
predicted to be almost the tri-bimaximal mixing.
The size of the extra dimension has a connection to the possible mass spectrum;
a small (large) volume corresponds to the normal (inverted) mass hierarchy.
\end{abstract}

\end{titlepage}

\newpage
\section{Introduction}
The origin of the fermion masses and mixing is one of the most compelling 
subjects in particle physics. In the past few decades, the neutrino oscillation 
experiments have revealed that the lepton mixing angles are quite different from 
the quark mixing matrix~\cite{nu}. 
While the off-diagonal components of the quark mixing matrix 
are much less than unity, the lepton mixing angles are excellently modeled by 
the tri-bimaximal mixing~\cite{TBM} containing two large and one vanishingly small 
angles. The observed generation structure has stimulated model building activities
toward the theory beyond the Standard Model (SM).

A phenomenological approach to explain the observed generation structure is to assume
symmetry among three generations of fermion.
Recently, discrete groups such as $S_3$~\cite{S3}, $A_4$~\cite{A4}, and $S_4$~\cite{S4}
have attracted much attention as feasible candidates for flavor symmetry in the 
lepton and quark sector.
A common issue among these models is how to break flavor symmetry
so that the observed data is naturally produced.
Probably the most popular mechanism for flavor symmetry breaking is the vacuum expectation
values (VEVs) of new scalar fields (flavons). 
Within this option, however, one is often forced to sacrifice the simplicity of the 
models due to multitude of flavons needed to construct desirable mass matrices.

Another mechanism available in the literature is to utilize twisted boundary 
conditions~\cite{SS} for bulk fields in extra spacial dimensions.
This idea is explored within the seesaw mechanism in five dimensions where
the $S_3$ permutation symmetry is broken by the boundary conditions for the
bulk right-handed neutrinos~\cite{TF}. 
In this scheme, symmetry-breaking effects are calculable and
the tri-bimaximal mixing is obtained as a consequence of symmetry breaking
prompted by many possible boundary conditions.
Furthermore, extra dimensions help VEVs of the flavons to be aligned so that 
viable neutrino mass matrices are obtained~\cite{TFS}.
Since theories with extra dimensions have attracted great attention
as a feasible paradigm to understand the gauge hierarchy problem~\cite{ADD,RS},
the interplay between flavor symmetry and extra dimensions, it is an intriguing 
subject to investigate.

In this paper, we discuss the neutrino model based on the five-dimensional framework 
of Ref.~\cite{TF}.
The novel point in this work is the adoption of the $S_4$ symmetry as an alternative
to $S_3$.
Although the $S_4$ group also belongs to the symmetric group as $S_3$,
it differs from $S_3$ in many ways, so that the whole picture of the model 
and the physical consequences are significantly different from what is found in 
the previous study.
For example, the $S_4$ group has various irreducible representations; two singlets,
one doublet, and two triplets.
By identifying the three generations of fermions as triplets, the number of 
the Yukawa couplings is restricted to be small, so that the model
prediction becomes rich and testable at future experiments.
In addition, the variety of the irreducible representation makes it possible
to construct realistic charged-lepton mass matrices.
Another point is the order of $S_4$ ($4! = 24$) which is much larger than 
$S_3$ ($3! = 6$). With the 24 group elements in $S_4$, the number of the
theoretically possible boundary conditions for the bulk fermions is large 
($100$ patterns) and the theory contains new types of the lepton mass matrices
which  are unavailable in the $S_3$ model.

This paper is organized as follows.
In Section~2, we discuss the basic setup of the model;
general boundary conditions for the bulk neutrinos, 
locations of the SM fields, the Kaluza-Klein (KK) expansion etc. 
In Section~3, we present two concrete models of the $S_4$ breaking and 
discuss the predictions which can be used as markers in future neutrino experiments.
Section~4 is devoted to summarizing the results.
Appendices~\ref{LG}  and~\ref{s4g} summarize our convention for Lorentz spinors and
some basic properties of the $S_4$ group, respectively.

\medskip
\section{Basic framework}
\label{setup}

We consider the fifth spacial dimension $y$ compactified on the $S^1/Z_2$ orbifold
with a radius $R$. There exist two fixed points at $y =0$ and $\pi R$. 
The four-component bulk fermions $\Psi_i(x,y)$ ($i=1,2,3$) are introduced as 
three generations of the right-handed neutrinos and their chiral partners~\cite{bulk_nuR}.
There are two boundary conditions linked to the two operations on the $S^1/Z_2$ 
space; the translation $\hat{T} : y \to y + 2\pi R$ and the parity 
$\hat{Z}: y \to -y$. The boundary conditions for the bulk fermions $\Psi_i(x,y)$ 
are written as
\begin{eqnarray}
\Psi_i(x,-y) = Z_{ij}\otimes\gamma_5 \Psi_j(x,y), \quad\quad 
\Psi_i(x,y + 2\pi R) = T_{ij} \, \Psi_j(x,y), 
\label{zt}
\end{eqnarray}
where $Z_{ij}$ and $T_{ij}$ are the matrices acting on the generation space. 
The parity and translation imply that $Z^2 = 1$ and $ZT = T^{-1}Z$ must
be satisfied. 
Instead of the translation $T_{ij}$ in (\ref{zt}), one may use another
parity $Z'=TZ$ to write the boundary conditions 
\begin{eqnarray}
\Psi_i(x,-y) = Z_{ij}\otimes\gamma_5 \Psi_j(x,y), \qquad
\Psi_i(x,\pi R - y) = Z'_{ij}\otimes\gamma_5 \Psi_j(x,\pi R + y). 
\label{zzp}
\end{eqnarray}
The parity $Z'_{ij}$ is the reflection with respect to $y = \pi R$. 
In this case, the two matrices must obey $Z^2 = 1$ and $Z'{}^2 = 1$
as consistency relations.

The SM fields including the left-handed lepton doublets $L_i = P_L L_i
= \big(\begin{smallmatrix} \nu_{L_i} \\ e_{L_i}^{}\end{smallmatrix}\big)$
and the Higgs field $H$ are assumed to be localized at $y = \pi R$.
This SM-field profile gives an example
and the analysis below is applied to other cases in a similar way.
The Lagrangian including the neutrino fields is given by
\begin{eqnarray}
\mathcal{L} &=&
i\overline{\Psi}_j \Gamma^M \partial_M \Psi_j
\,-\, \frac{1}{2} \big( \overline{\Psi^c_i} (M_{ij})\Psi_j + {\rm h.c.}\big)
\nonumber \\
&&
-\frac{1}{\sqrt{\Lambda}}\left(\,  \overline{\Psi}_i(Y_{\nu_{ij}}) L_j H 
\,+\,  \overline{\Psi^c_i} (Y_{\nu_{ij}}^c) L_j H
\,+\, {\rm h.c.}\right) \delta(y - \pi R),
\label{L}
\end{eqnarray}
where $\Lambda$ stands for the fundamental scale of the theory. 
In this paper, we do not consider Lorentz-violating mass terms
such as $\overline{\Psi}\Gamma^5\Psi$, $\overline{\Psi^c}\Gamma^5\Psi$~\cite{Gamma5M}
and the bulk Dirac (kink) mass while they are often discussed in 
literature. After the electroweak symmetry breaking, 
the boundary interactions develop the neutrino Dirac masses; $m_{ij} = Y_{\nu_{ij}}v$ 
and $m^c_{ij} = Y_{\nu_{ij}}^c v$ where $v$ is
the VEV of the Higgs field $\langle H \rangle = 
\big(\begin{smallmatrix}0 \\ v\end{smallmatrix}\big)$. 
The charge-conjugate spinor $\Psi^c$ is defined 
by $\Psi^c \equiv \Gamma^3\Gamma^1 \overline{\Psi}^{\rm T}$. Our
convention for the gamma matrices and Lorentz spinors are given in
Appendix \ref{LG}.

The four-dimensional effective theory is described by the KK modes
of the right-handed neutrinos.
Under a set of boundary conditions, the bulk fermions $\Psi_i(x,y)$ are
expanded as
\begin{equation}
\Psi_i(x,y) \;=\; \left(\begin{array}{c}
\displaystyle
\sum_{n=0}^\infty \chi_{R_{ij}}^n(y) \psi_{R_j}^n(x) \\[2mm]
\displaystyle
\sum_{n=0}^\infty \chi_{L_{ij}}^n(y) \psi_{L_j}^n(x)
\end{array}\right)
\label{KKex}
\end{equation}
with the orthogonal systems $\chi_{R,L}^n(y)$. 
It is convenient to choose them to satisfy the normalization conditions
$
\int_0^{\pi R}\!\! dy \,\, \Big[\, \chi_{R,L}^m{}^\dag 
 \chi_{R,L}^n{} \,\Big]_{ij} = \delta_{mn}\delta_{ij}
$
so that the kinetic term of each KK mode is canonically normalized.
By substituting the expansion into the five-dimensional Lagrangian
and integrating it over the extra space, we have the four-dimensional
effective Lagrangian
\begin{eqnarray}
\mathcal{L}_4 \,=\, i N^\dag  \sigma^\mu \partial_\mu N
- \frac{1}{2}\left( N^{\rm T} \epsilon \otimes M_N N
 + {\rm h.c.} \right),
\label{L4}
\end{eqnarray}
\begin{eqnarray}
M_N  = \!
\left( \begin{array}{c|c}
 0 &  \,\, M_D^{\rm T} \, \\[1mm]\hline
 &  \\
\!\!M_D\! &  \;\; M_H  \, \\
 &  \\
\end{array}\right)
=
\left( \begin{array}{c|ccccc}
 & M_0^{\rm T} & M_0^c{}^{\rm T} & M_1^{\rm T} & M_1^c{}^{\rm T} & \cdots  \\\hline
M_0 & -M_{R_{00}}^* &  M_{K_{00}} &  &  &  \cdots \\
M_0^c &  M_{K_{00}}^{\rm T} & M_{L_{00}} &   &  & \cdots \\
M_1 &  &  & -M_{R_{11}}^* & M_{K_{11}} &  \cdots \\
M_1^c &  & &   M_{K_{11}}^{\rm T} &  M_{L_{11}} & \cdots \\
\vdots & \vdots & \vdots & \vdots & \vdots & \ddots  \\
\end{array}
\right)
\!\!,\,\,
N =\!
\begin{pmatrix}
\nu_L^{} \\[.5mm]  \epsilon\psi_R^0{}^* \\[.5mm]
 \psi_L^0 \\
\epsilon \psi_R^1{}^* \\[.5mm]  \psi_L^1 \\
\vdots \\
\end{pmatrix}\!\!,\;\;
\label{KKmm}
\end{eqnarray}
where $\epsilon$ is the antisymmetric tensor and
\begin{gather}
M_{K_{mn}} \,=\, \int_0^{\pi R}\!\!\! dy \,\, \chi_R^m{}^\dag 
(-\partial_y) \chi_L^n , \nonumber\\
M_{R_{mn}} \,=\,
 \int_0^{\pi R}\!\!\! dy  \,\, \chi_R^m{}^{\rm T} M \chi_R^n, \quad
M_{L_{mn}} \,=\, \int_0^{\pi R}\!\!\! dy \,\,
\chi_L^m{}^{\rm T} M \chi_L^n, \nonumber\\
M_n  \,=\,  \frac{1}{\sqrt{\Lambda}}\chi_R^n{}^\dag(\pi R)m,
\quad
M_n^c  \,=\,  \frac{1}{\sqrt{\Lambda}}\chi_L^n{}^{\rm T}(\pi R) m^c.
\label{masses0}
\end{gather}
The zero modes are suitably subtracted according to the boundary conditions.
The generation indices in~(\ref{L4}),~(\ref{KKmm}),~(\ref{masses0}) are suppressed 
for simplicity. 
We leave the entries connecting different KK levels blank since these entries
will be vanishing in the following discussion\footnote{This does not hold in 
more general/different setups, e.g., on curved background or the boundary 
Majorana mass~\cite{WY}.}. 

The mass spectrum of Majorana neutrinos is obtained by 
diagonalizing $M_N$. For $M_D \ll M_H$,
the seesaw mechanism is available and the Majorana mass matrix of the light 
neutrinos is approximately given by
\begin{eqnarray}
M_\nu \,=\, - M_D^{\rm T} M_H^{-1} M_D.
\label{seesaw}
\end{eqnarray}
In what follows, we assume that the seesaw mechanism works.
That is, the inverse of the compactification radius and/or the bulk Majorana
scale are much larger than the boundary Dirac masses $M_n$ and $M_n^c$.

\medskip

\section{$S_4$ symmetry breaking and its consequences}
\label{appli}
In this section, we introduce the $S_4$ discrete symmetry and calculate
its breaking effects on the neutrino mass matrix.
We present $S_4$ charge assignment and the possible boundary conditions allowed 
by the consistency conditions. 
Two particular boundary conditions are discussed in detail
as viable examples.
Finally we comment on possible constructions of the charged-lepton sector.

\subsection{Charge assignment and possible boundary conditions}
The irreducible representations of $S_4$ are two singlets $\underline{1}$
and $\underline{1}'$, one doublet $\underline{2}$, and two triplets 
$\underline{3}$ and $\underline{3}'$ (see Appendix~\ref{s4g}).
Suppose that the three generations of the bulk fermions  $\Psi_i(x,y)$ and 
the lepton doublets $L_i(x)$ behave as triplet $\underline{3}$.
The symmetry-invariant mass parameters in the five-dimensional 
Lagrangian~(\ref{L}) are then written as
\begin{eqnarray}
M_{ij} = M\delta_{ij}, \quad\,\,
m_{ij} = m\delta_{ij}, \quad\,\,
m^c_{ij} = m^c\delta_{ij}.
\end{eqnarray} 
If symmetry is preserved, the neutrino mass matrix $M_\nu$ 
is also proportional to the identity matrix, which is not consistent with
the observations.
In fact, the trivial boundary conditions $Z =1$ and $T =1$ lead to the
neutrino mass matrix~\cite{WY}
\begin{eqnarray}
M_{\nu_{ij}} = \frac{1}{\Lambda R} \, \frac{|M|R}{\tanh(\pi |M|R)} \, \frac{m^2}{M^*}
\delta_{ij}.
\label{s4s}
\end{eqnarray}
Symmetry breaking is the key to obtain the flavor mixture and the mass splittings 
between three generations.

It is convenient to specify the boundary conditions by $Z$ and $Z'$ with which 
the consistency relations are written as the parity conditions $Z^2 = 1$ and $Z'{}^2 = 1$.
\begin{table}[t]
\begin{center}
\begin{tabular}{cc}
 & $Z'$ \\
$Z$ & 
\begin{tabular}{|c||c|c|c|c|c|c|c|c|c|c|}\hline
 &$a_1$&$a_2$&$a_3$&$a_4$&$d_1$&$d_2$&$f_1$&$f_3$&$e_1$&$e_4$\\[.5mm]\hline\hline
$a_1$&~~~~~&~~~~~&~~~~~&~~~~~~&~~~~~&~~~~~&~~~~~&~~~~~&~~~~~&~~~~~ \\[.5mm]\hline
$a_2$&&&&&&&${\cal C}_{1}$&${\cal C}_{2}$&${\cal C}_{3}$&${\cal C}_{4}$ \\[.5mm]\hline
$a_3$&&&&&${\cal C}_{5}$&${\cal C}_{6}$&&&${\cal C}_7$&${\cal C}_8$ \\[.5mm]\hline
$a_4$&&&&&${\cal C}_{9}$&${\cal C}_{10}$&${\cal C}_{11}$&${\cal C}_{12}$&& \\[.5mm]\hline
$d_1$&&&${\cal B}_{1}$&${\cal B}_{2}$&&&${\cal A}_1$&${\cal A}_{2}$&${\cal A}_{3}$&${\cal A}_{4}$\\[.5mm]\hline
$d_2$&&&${\cal B}_{3}$&${\cal B}_{4}$&&&${\cal A}_5$&${\cal A}_6$&${\cal A}_{7}$&${\cal A}_{8}$\\[.5mm]\hline
$f_1$&&${\cal B}_{5}$&&${\cal B}_6$&${\cal A}_{9}$&${\cal A}_{10}$&&&${\cal A}_{11}$&${\cal A}_{12}$\\[.5mm]\hline
$f_3$&&${\cal B}_{7}$&&${\cal B}_8$&${\cal A}_{13}$&${\cal A}_{14}$&&&${\cal A}_{15}$&${\cal A}_{16}$\\[.5mm]\hline
$e_1$&&${\cal B}_{9}$&${\cal B}_{10}$&&${\cal A}_{17}$&${\cal A}_{18}$&${\cal A}_{19}$&${\cal A}_{20}$&&\\[.5mm]\hline
$e_4$&&${\cal B}_{11}$&${\cal B}_{12}$&&${\cal A}_{21}$&${\cal A}_{22}$&${\cal A}_{23}$&${\cal A}_{24}$&&\\[.5mm]\hline
\end{tabular}\\
\end{tabular}
\end{center}
\caption{The possible boundary conditions in terms of the two parities $Z$ and $Z'$.
The symbols $a_1, a_2, a_3, \cdots, e_4$ stand for the group elements (see 
Appendix~\ref{s4g}).
The calligraphic characters represent the boundary conditions with which the
theory becomes viable for neutrino phenomenology.
Out of the 100 general possibilities, 48 patterns are useful.}
\label{t1}
\end{table}
Table~\ref{t1} shows the possible combinations of $Z$ and $Z'$.
In the $S_4$ group, there are 10 elements satisfying 
the parity condition, 
so that there are total 100 possibilities in the table.
The boundary conditions are classified into 14 categories according to their 
physical consequences. 
Out of the 14 categories of the boundary condition,
the three categories tagged by the calligraphic characters 
${\cal A}, {\cal B}, {\cal C}$ prompt $S_4$ breaking which is viable for neutrino 
phenomenology.
The subscripts are the serial numbers in each category.  
The conditions in the same category produce identical neutrino mass matrices up to 
the rotations by the group elements.
For instance, the seesaw induced mass $M_\nu$ for the condition ${\cal B}_1$ is
obtained by exchanging the second and third  generation labels of ${\cal B}_2$,
and vice versa.
Since such rotations are absorbed by appropriate $S_4$ transformations of 
the left-handed neutrinos, physical implications in each category are equivalent.

Within the three useful categories of the boundary conditions, the $S_4$ symmetry 
is completely broken and the mass matrix at the low-energy acquires structure 
sufficiently rich to account for the observed neutrino masses and mixing.
Since the neutrino mass matrix with the tri-bimaximal mixing is invariant under 
the $S_2$ transformations, one may naively expect that $S_4 \to S_2$ breaking
patterns are feasible for neutrino phenomenology.
However, such categories produce too simple mass matrices to be realistic.
For example, many  $S_4 \to S_2$ breaking conditions predict
that at least two neutrino masses are degenerate.
In what follows, we discuss physical implications of 
the three categories ${\cal A}, {\cal B}$ and ${\cal C}$ 
in detail.

\subsection{Model I: The category ${\cal A}$}
\label{moA}
Let us first discuss the category ${\cal A}$.
Out of the 24 patterns in ${\cal A}$, we focus on 
${\cal A}_9$:
\begin{eqnarray}
Z = f_1 = \begin{pmatrix}
0&0&1\\ 0&1&0 \\ 1&0&0 \\
\end{pmatrix}
,\quad\quad
Z' = d_1 = \begin{pmatrix}
1&0&0 \\ 0&0&1 \\ 0&1&0 \\
\end{pmatrix},
\end{eqnarray}
as a concrete example which is convenient for presentation.
The KK expansion~(\ref{KKex}) which satisfies the boundary conditions is given by
\begin{eqnarray}
&&\chi_R^0(y) = \frac{1}{\sqrt{\pi R}}V
\begin{pmatrix}
\frac{1}{\sqrt{2}}e^{i \frac{y}{3R}}
&0&0 \\
\frac{1}{\sqrt{2}}e^{-i \frac{y}{3R}}
&0&0 \\
0&0& 1 \\
\end{pmatrix},\quad
\chi_L^0(y) = \frac{1}{\sqrt{\pi R}}V
\begin{pmatrix}
\frac{1}{\sqrt{2}}e^{i\frac{y}{3R}}&0&0 \\
\frac{-1}{\sqrt{2}}e^{-i\frac{y}{3R}}&0&0 \\
0&0&0  \\
\end{pmatrix},
\nonumber\\
&&\chi_R^n(y) = \sqrt{\frac{2}{\pi R}}V
\begin{pmatrix}
\frac{1}{2}e^{i\left( n + \frac{1}{3}\right)\frac{y}{R}}
& \frac{1}{2}e^{-i\left( n - \frac{1}{3}\right)\frac{y}{R}}&0 \\
\frac{1}{2}e^{-i\left( n + \frac{1}{3}\right)\frac{y}{R}}
& \frac{1}{2}e^{i\left( n - \frac{1}{3}\right)\frac{y}{R}}&0 \\
0&0& \cos\left( \frac{n}{R}y \right) \\
\end{pmatrix}\quad\quad ( n \geq 1), \nonumber\\
&&\chi_L^n(y) =  \sqrt{\frac{2}{\pi R}}V
\begin{pmatrix}
\frac{1}{2}e^{i\left( n + \frac{1}{3}\right)\frac{y}{R}}
& \frac{-1}{2}e^{-i\left( n - \frac{1}{3}\right)\frac{y}{R}}&0 \\
\frac{-1}{2}e^{-i\left( n + \frac{1}{3}\right)\frac{y}{R}}
& \frac{1}{2}e^{i\left( n - \frac{1}{3}\right)\frac{y}{R}}&0 \\
0&0& \sin\left( \frac{n}{R}y \right) 
\label{type3}
\end{pmatrix}\quad\quad ( n \geq 1),
\end{eqnarray}
where $V$ is the unitary matrix
\begin{eqnarray}
V  = 
\frac{1}{\sqrt{3}}
\begin{pmatrix}
\omega & \omega^2 & 1 \\
1 & 1& 1& \\
\omega^2 & \omega & 1 \\
\end{pmatrix}
\label{U}
\end{eqnarray}
with $\omega \equiv e^{i2\pi /3}$.
Three ``zero modes'' are absent in this boundary condition
(here $\psi^0_{R_2}, \psi^0_{L_2},\psi^0_{L_3}$ are taken as such absent fields).
In order to satisfy the boundary condition, nontrivial generation structure is 
necessary in the KK wave functions.

By substituting these KK expansions into~(\ref{masses0}) and
performing the seesaw operation~(\ref{seesaw}), one finds the Majorana mass matrix 
\begin{eqnarray}
M_\nu  &=& 
\frac{1}{\Lambda R} \left[\, 
\frac{s |M|R }{c + 1/2}
\frac{m^2}{M^*}
\begin{pmatrix}
\frac{4}{6}&\frac{-2}{6}&\frac{-2}{6}\\\frac{-2}{6}&\frac{1}{6}&\frac{1}{6}\\
\frac{-2}{6}&\frac{1}{6}&\frac{1}{6}\\
\end{pmatrix}
+ 
\frac{|M|R}{\tanh(\pi |M|R)}\frac{m^2}{M^*}
\begin{pmatrix}
\frac{1}{3}&\frac{1}{3}&\frac{1}{3} \\\frac{1}{3}&\frac{1}{3}&\frac{1}{3} 
\\\frac{1}{3}&\frac{1}{3}&\frac{1}{3}\\
\end{pmatrix} \right. \nonumber\\
&& \quad\quad \left. 
-\frac{s |M|R }{c + 1/2}
\frac{(m^c)^2}{M}
\begin{pmatrix}
&&\\&\frac{1}{2}&\frac{-1}{2}\\ &\frac{-1}{2}&\frac{1}{2} \\
\end{pmatrix}
-
\frac{|M|R}{c + 1/2}
\frac{m m^c}{|M|}
\begin{pmatrix}
&\frac{1}{2}&\frac{-1}{2}\\\frac{1}{2}&\frac{-1}{2}&\\
\frac{-1}{2} &&\frac{1}{2} \\
\end{pmatrix} \,\right],
\label{ModelA}
\end{eqnarray}
where $c \equiv \cosh(2\pi |M|R)$ and $s \equiv \sinh(2\pi |M|R)$.
It is noticed that the $S_4$ symmetry is completely broken in~(\ref{ModelA}).
However, in the limit $|M|R \gg 1$, the last term becomes negligible and 
$M_\nu$ recovers the $S_2$ symmetry~($2\leftrightarrow 3$ exchange).
This is because the theory has two breaking sources at $y=0$ and $y=\pi R$, 
and $S_4$ is entirely broken only globally.
In the large-size limit of extra dimension $|M|R \gg 1$,
the boundary condition at the distant brane ($f_1$ at $y=0$) becomes irrelevant 
to physics at another boundary, whereas the local
twisting ($d_1$ at $y=\pi R$) which respects the 2-3 exchange symmetry 
remains relevant.

If the last term of~(\ref{ModelA}) is vanishing, 
the Majorana mass matrix~(\ref{ModelA}) is diagonalized by
the tri-bimaximal mixing~\cite{TBM} 
\begin{equation}
V_\text{tri-bi} = 
\begin{pmatrix}
\frac{2}{\sqrt{6}} &  \frac{1}{\sqrt{3}} & 0 \\
\frac{-1}{\sqrt{6}} & \frac{1}{\sqrt{3}} &  \frac{-1}{\sqrt{2}} \\
 \frac{-1}{\sqrt{6}} &  \frac{1}{\sqrt{3}} &   \frac{1}{\sqrt{2}}
\end{pmatrix}.
\label{tri-bi}
\end{equation}
Besides the large-volume limit mentioned above, that is also realized by taking 
$m^c =0$ which predicts the inverted hierarchy with 
$m_3 = 0$. For the normal hierarchy, both $m \neq 0$ and $m^c \neq 0$ are necessary, 
so that the neutrino mixing is deviated from the tri-bimaximal form by presence of 
the last term.
The deviation induces a nonzero reactor angle $\theta_{13}$ sufficiently
large to be measurable in forthcoming experiments.

Let us examine the mass matrix~(\ref{ModelA}) and its predictions in detail. 
It is useful to rewrite the matrix~(\ref{ModelA}) as
\begin{eqnarray}
 M_\nu=
\frac{-|M|}{\Lambda }
V_\text{tri-bi}
 \begin{pmatrix}
\frac{-2s}{2c+1}
\frac{m^2}{M^*}  & 0 
& \frac{\sqrt3}{2c+1}
\frac{m m^c}{|M|} \\ 
                   0    & 
\frac{-1}{\tanh( \pi |M|R)}
\frac{m^2}{M^*}   &0    \\
       \frac{\sqrt3}{2c+1}
\frac{m m^c}{|M|}  & 0  
& \frac{2s}{2c+1}
\frac{(m^c)^2}{M}   \\
 \end{pmatrix}
V_\text{tri-bi}^{\rm T}.
\label{massmatrixA}
\end{eqnarray}
The neutrino masses are given by
\begin{eqnarray}
\begin{split}
m_1
&= \frac{|m|^2}{\Lambda}\frac{1}{2c+1} 
\left| \, s(1-r^2)+\sqrt{s^2(1+r^2)^2+3r^2} \,\right|,
\\
m_2&= \frac{|m|^2}{\Lambda}\frac{1}{2c+1}\left[\, \frac{2c+1}{\tanh(\pi |M|R)} \,\right],
\\
m_3
&=\frac{|m|^2}{\Lambda}\frac{1}{2c+1} 
\left|\, s(1-r^2)-\sqrt{s^2(1+r^2)^2+3r^2}\,\right|,
\label{masses}
\end{split}
\end{eqnarray}
where $ r \equiv |m^c|/|m|$.
If the charged-lepton mass matrix is diagonal,
the lepton mixing matrix $U$ is identified as the unitary matrix
which diagonalizes~(\ref{massmatrixA});
\begin{equation}
U= V_\text{tri-bi} 
\begin{pmatrix}
e^{i\rho} & 0 & 0\\ 
0 & e^{i\rho}& 0   \\
0 & 0 & e^{i\sigma}   \\
\end{pmatrix} 
\begin{pmatrix}
\cos\theta&0&\sin\theta \\ 
0&1&0 \\
-\sin\theta&0&\cos\theta \\
\end{pmatrix}
\begin{pmatrix}
1&0&0\\0&1&0\\0&0&i\\
\end{pmatrix},
\label{theta}
\end{equation}
where $\rho \equiv \arg(m)+\arg(M)/2$,  $\sigma \equiv \arg(m^c)-\arg(M)/2$ and
\begin{equation}
\tan 2\theta= \frac{\sqrt{3}r}{s(1+r^2)}.
\label{mixing}
\end{equation}
The relevant mixing matrix elements are written as 
\begin{eqnarray}
U_{e2}=\frac{1}{\sqrt{3}} e^{i\rho},\quad
U_{e3}=\frac{2i}{\sqrt{6}}\sin\theta e^{i\rho},\quad
U_{\mu_3}=-i \left( \frac{1}{\sqrt{2}}\cos\theta e^{i\sigma}
+ \frac{1}{\sqrt{6}}\sin\theta e^{i\rho} \right).
\label{MNS}
\end{eqnarray}
The solar angle is robustly predicted to be $\sin\theta_{12} = 1/\sqrt{3}$ 
up to the small corrections of $\mathcal{O}(\theta_{13}^2)$.
The mixing matrix $U$ contains the phase parameters which cannot be removed
by field redefinitions.
That is, the boundary condition induces not only the $S_4$ breaking but also
CP violation.
The mass formulas~(\ref{masses}) involve three effective parameters; 
$|m|^2/\Lambda$, $|M|R$ and $r$.
These parameters are determined if the three neutrino masses are
regarded as fixed observables.

The neutrino mass matrix~(\ref{massmatrixA}) accommodates all possible
mass patterns of neutrinos; normal, inverted and degenerate.
The mass pattern is determined by the mass ratio $r$.
In the region where $r>1 \,(r<1)$, the normal (inverted) mass ordering 
is realized. In the case of $r=1$, neutrino masses are degenerate
since $|m_1|=|m_3|$.

At first, let us discuss the normal mass ordering ($r>1$).
From the mass formulas~(\ref{masses}),
it is seen that a hierarchical spectrum is realized in the regime that 
$r \gg 1$, $|M|R \ll 1$ and $|M|Rr > \sqrt{3}/2\pi$.
In such a regime, the ratio of the two mass squared differences 
is approximated as
\begin{eqnarray}
\frac{m_3^2-m_1^2}{m_2^2-m_1^2}
\,\simeq\,
\frac{8x^3\sqrt{3+4x^2}}
{9-3x^2-8x^4+4x^3\sqrt{3+4x^2}},
\end{eqnarray}
where $x= \pi |M|Rr$. By substituting the typical  observed value
  $|\Delta m^2_{31}|/\Delta m^2_{21}\simeq 30$ into the left-hand side,
one obtains $x\approx 1.95$. 
The reactor angle is then predicted as
\begin{eqnarray}
\sin\theta_{13} \simeq \frac{r}{\sqrt{2}s(1+r^2)}\simeq
\frac{1}{2\sqrt{2} x}
\frac{r^2}{1+r^2}
\simeq \frac{1}{2\sqrt{2} x}
\simeq 0.18.
\label{Ue3}
\end{eqnarray}
\begin{figure}
\begin{center}
\scalebox{0.85}{
\includegraphics{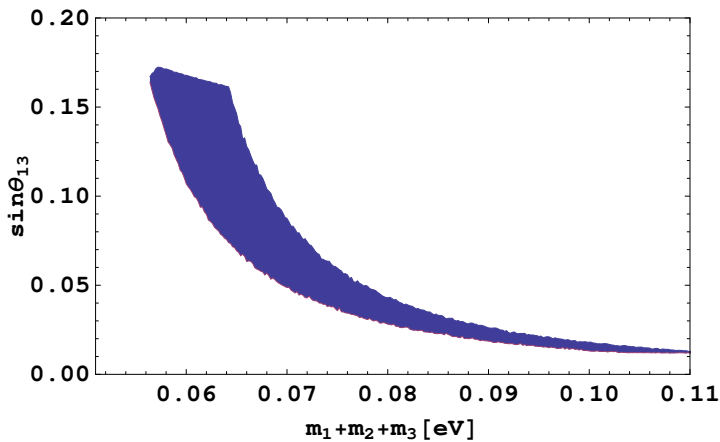}
}
\caption{The reactor angle $\sin\theta_{13}$ as a function of $m_1+m_2+m_3$ 
for the normal mass ordering in the Model~I.}
\label{fig1}
\end{center}
\end{figure}
Interestingly, the reactor angle is predicted to be just below the current upper bound.
The discovery of $\theta_{13}$ is imminent if the normal hierarchy is realized in 
this model. 
It is seen in~(\ref{Ue3}) that $\theta_{13}$ decreases as the parameter $r$
decreases, which means that $\theta_{13}$ takes maximum value at the hierarchical
limit and it is lowered as the spectrum is shifted to the degenerate pattern. 

Figure~\ref{fig1} shows the numerical plot of $\sin\theta_{13}$ as a function
of the total neutrino mass $\Sigma \equiv \sum_i m_i$.
As seen in~(\ref{masses}) and~(\ref{mixing}), the three parameters
$|m|^2/\Lambda$, $|M|R$ and $r$ are determined by regarding $|\Delta m_{31}^2|$, 
$\Delta m_{21}^2$ and $\Sigma$ as inputs. 
That is, $\sin\theta_{13}$ is plotted as a function of $\Sigma$ with fixed
values of $|\Delta m_{31}^2|$ and $\Delta m_{21}^2$.
In the numerical calculation, the $3\sigma$ ranges of the mass squared
differences in Ref.~\cite{3nu} are used.
The cutoff behavior of~(\ref{Ue3}) is seen at the lower end of $\Sigma$,
i.e., at the hierarchical limit.
In the hierarchical regime that $\sqrt{|\Delta m^2_{31}|} \lesssim \Sigma \lesssim 0.06 
\, {\rm eV}$, $ 0.07 \lesssim \sin\theta_{13} \lesssim 0.17$, which is expected to be 
observed at T2K, Double Chooz, RENO and Daya Bay~\cite{ue3experiments}.
As $\Sigma$ increases, that is, the three neutrino masses approach 
the degenerate pattern, $\sin\theta_{13}$ becomes smaller.
This behavior is also understood in view of~(\ref{massmatrixA}). 
It is seen that the three diagonal components get closer to each other
in the regime $|M|R\gg 1$ and $r\simeq 1$.
The off-diagonal elements are then suppressed by a large $\cosh(2\pi |M|R)$ factor
compared to the diagonal ones.

\begin{figure}
\begin{center}
\scalebox{0.85}{
\includegraphics{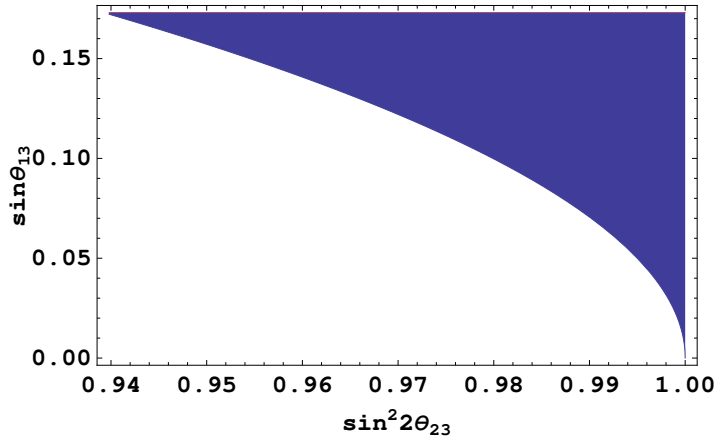}
}
\caption{Correlation between $\sin^2 2\theta_{23}$ and $\sin\theta_{13}$
for the normal mass hierarchy in the Model~I.}
\label{fig2}
\end{center}
\end{figure}

The atmospheric angle $\theta_{23}$ is, as seen in~(\ref{MNS}), correlated to 
$\theta_{13}$.
If CP is conserved, $\theta_{23}$ is deviated from the maximum 
in such a way that $\sin\theta_{23} \simeq 1/\sqrt{2} + 
\sin\theta_{13}/2$.
With a finite combination of the phase parameters $\rho-\sigma$, however, 
the effect of $\theta_{13}$ can be canceled and the maximal $\theta_{23}$ is
possible even with a nonzero value of $\theta_{13}$.
Figure~\ref{fig2} presents the correlation between $\sin^2 2\theta_{23}$ and 
$\sin \theta_{13}$ where the shaded region shows the possible parameter space.
The lower bound of $\theta_{13}$ is predicted for each fixed value of
$\sin^2 2\theta_{23}$. 
In addition to the correlation between $\sin\theta_{13}$ and the
mass pattern, this will be also a crucial test of this model 
if $\sin^2 2\theta_{23}$ will be  precisely determined at the T2K experiment.

Next let us discuss the inverted mass ordering ($r<1$). 
As seen in~(\ref{massmatrixA}) and~(\ref{masses}), 
$m_1$ and $m_2$ are close to each other if $|M|R\gg 1$.
In this regime, $\tanh(\pi|M|R) \simeq 1 - 1/s$ and
the ratio of the neutrino mass squared differences is given by
\begin{align}
\frac{m_1^2-m_3^2}{m_2^2-m_1^2} \,\simeq\, \frac{s(1-r^4)}{3}\ .
\end{align}
By using this relation, the reactor angle $\sin\theta_{13}$ is written by
$r$ and the mass differences;
\begin{equation}
\sin\theta_{13} \,\simeq\,
\frac{r}{\sqrt 2s(1+r^2)}\,\simeq\, \frac{r(1-r^2)}{3\sqrt{2}}
\frac{\Delta m^2_{21}}{|\Delta m^2_{31}|}.
\end{equation}
Even at the extremum $r=1/\sqrt 3$, $\sin\theta_{13} = \mathcal{O}(10^{-3})$ 
which is far below the expected sensitivity in future experiments.
After all, the reactor angle follows
\begin{equation}
\sin\theta_{13} \,\lesssim\, 0.003
\end{equation}
with typical values of the mass differences.
The lepton mixing  matrix is thus almost the tri-bimaximal mixing in the case of 
the inverted mass hierarchy.

\subsection{Model I\!I: The categories ${\cal B}$ and ${\cal C}$}
\label{moB}
Another interesting example is the categories ${\cal B}$ and ${\cal C}$. 
These two categories lead to quite similar physical consequences.
The predictions differ only in the solar angle $\theta_{12}$.
Hence we regard ${\cal B}$ and ${\cal C}$ as a single model and 
discuss ${\cal B}$ in detail.
We demonstrate the viability of the model with 
a concrete example ${\cal B}_9$:
\begin{eqnarray}
Z = e_1 = \begin{pmatrix}
0&1&0\\ 1&0&0 \\ 0&0&1 \\
\end{pmatrix},\quad\quad
Z' = a_2 = \begin{pmatrix}
1&0&0 \\ 0&-1&0 \\ 0&0&-1 \\
\end{pmatrix}.
\end{eqnarray}
The analysis below is applied to the other conditions in a similar manner. 
The KK expansion which satisfies ${\cal B}_9$ is given by
\begin{eqnarray}
&&\!\!\!\!\!\!\!\!\chi_R^0(y) = \frac{1}{\sqrt{\pi R}}\,
O\!\begin{pmatrix}
\cos\left(\frac{y}{4R}\right) &0&0 \\[1.5mm]
-\sin\left(\frac{y}{4R}\right)&0&0  \\
0&0&\sqrt{2}\cos\left(\frac{y}{2R}\right) \\
\end{pmatrix}, \label{k1}\\
&&\!\!\!\!\!\!\!\!\chi_R^n(y) = \frac{1}{\sqrt{\pi R}}\,
O\!
\begin{pmatrix}
\cos\!\left[ \left(n \!-\! \frac{1}{4} \right)\!\frac{y}{R} \right] &  
\cos\!\left[ \left(n \!+\! \frac{1}{4} \right)\!\frac{y}{R} \right] & 0 \\[1.5mm]
\sin\!\left[ \left(n \!-\! \frac{1}{4} \right)\!\frac{y}{R} \right] &  
-\sin\!\left[ \left(n \!+\! \frac{1}{4} \right)\!\frac{y}{R} \right] & 0 \\
0&0& \!\!\!\!\sqrt{2}\cos\!\left[ \left(n \!+\! \frac{1}{2} \right)\!
\frac{y}{R} \right] \\
\end{pmatrix}\!\,\, (n \geq 1),\\
&&\!\!\!\!\!\!\!\!\chi_L^0(y) =\frac{1}{\sqrt{\pi R}}\,
O\!
\begin{pmatrix}
\sin\left(\frac{y}{4R}\right) &0&0 \\[1.5mm]
\cos\left(\frac{y}{4R}\right) &0&0 \\
0&0&\sqrt{2}\sin\left(\frac{y}{2R}\right) \\
\end{pmatrix},\\
&&\!\!\!\!\!\!\!\!\chi_L^n(y) = \frac{1}{\sqrt{\pi R}}\,
O\!
\begin{pmatrix}
\sin\!\left[ \left(n \!+\! \frac{1}{4} \right)\! \frac{y}{R} \right] &  
\sin\!\left[ \left(n \!-\! \frac{1}{4} \right)\! \frac{y}{R} \right] &0  \\[1.5mm]
\cos\!\left[ \left(n \!+\! \frac{1}{4} \right)\! \frac{y}{R} \right] &  
-\cos\!\left[ \left(n \!-\! \frac{1}{4} \right)\!\frac{y}{R} \right] &0  \\
0&0& \!\!\!\!\!\sqrt{2}\sin\left[ \left(n \!+\! \frac{1}{2} \right)\!
\frac{y}{R} \right] \\
\end{pmatrix}\! \,\, (n \geq 1),\label{k4}
\end{eqnarray}
where $O$ is the orthogonal matrix
\begin{eqnarray}
O \,=\, \begin{pmatrix}
\frac{1}{\sqrt{2}}& \frac{-1}{\sqrt{2}} & 0 \\
\frac{1}{\sqrt{2}}& \frac{1}{\sqrt{2}} & 0 \\
0&0&1 \\
\end{pmatrix}.
\end{eqnarray}
Two ``zero modes'' are absent under this boundary condition
(here $\psi^0_{R_2}, \psi^0_{L_2}$ are taken as such absent fields).
With the KK functions~(\ref{k1})-(\ref{k4}),
the neutrino mass matrix at low energy turns out
\begin{eqnarray}
 M_\nu=
-\frac{|M|}{\Lambda}\tanh(2\pi |M|R)
\left(\begin{array}{ccc}
\frac{(m^c)^2}{M} & \frac{1}{s}\frac{m m^c}{|M|} & 0 \\ 
 \frac{1}{s}\frac{m m^c}{|M|} & -\frac{m^2}{M^*}  &0    \\[1.5mm]
 0  & 0  &  \frac{\tanh(\pi |M|R)}{\tanh(2\pi |M|R)}\frac{(m^c)^2}{M}\\
 \end{array}\right).
\label{massmatrixB}
\end{eqnarray}
In this model, the solar angle $\theta_{12}$ originates in the neutrino sector
while the reactor and atmospheric angles are generated in the charged-lepton 
sector. A suitable charged-lepton mass matrix will be discussed in the next 
subsection. For the category~${\cal C}$, the mass matrix~(\ref{massmatrixB})
is modified by the $45^\circ$ rotation for the 1-2 sector.

By diagonalizing the upper-left $2\times 2$ submatrix in~(\ref{massmatrixB}), 
one finds the mass eigenvalues and the mixing angle
\begin{eqnarray}
&&m_\pm \,=\,\left|\, 
\frac{r^2-1}{2}\pm\frac12\sqrt{\frac{4r^2}{s^2}+(r^2+1)^2}\,\right|
\frac{|m|^2\tanh(2\pi |M|R)}{\Lambda}\ , \nonumber \\
&&m_3 \,=\, \left  (\frac{\tanh({\pi |M|R})}{\tanh({2\pi |M|R})}r^2 \right )
\frac{|m|^2\tanh(2\pi |M|R)}{\Lambda} \ , \nonumber \\
&&\tan2\theta_{12} \,=\,\frac{2r}{s(r^2+1)}.
\label{massesB}
\end{eqnarray}
\begin{figure}
\begin{center}
\scalebox{0.85}{
\includegraphics{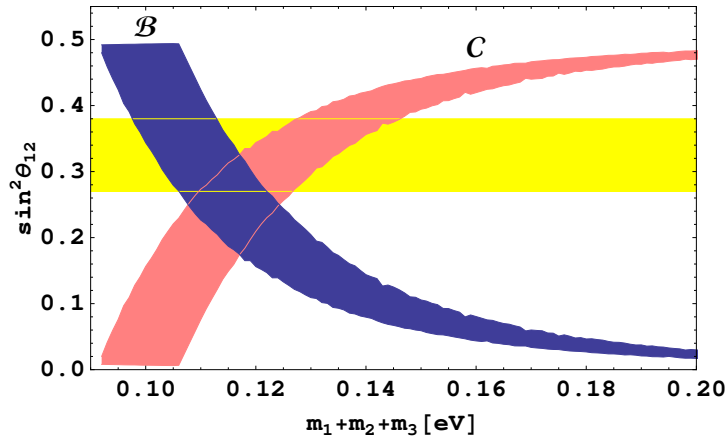}
}
\caption{The solar angle $\sin^2 \theta_{12}$ as a function of $m_1+m_2+m_3$ 
in the Model~I\!I. The horizontal lines shows the $3\sigma$ allowed range of
the solar angle~\cite{3nu}.}
\label{fig3}
\end{center}
\end{figure}
This model accommodates only the inverted mass ordering since
$m_3$ cannot be largest among the three eigenvalues.
The two eigenvalues $m_\pm$ are identified as $m_2=m_{+(-)}$ and 
$m_1=m_{-(+)}$ for $r>1 \ (r<1)$ and degenerate for $r=1$.
The observed mass differences are not reproduced in the regime that 
$|M|R \gtrsim 1/2\pi$.
In such a regime, the ratio of the mass differences is given by
$|\Delta m^2_{31}|/\Delta m^2_{21} \simeq -1$ for $r>1$ and
$2r^4/s$ for $r<1$, which contradict the data.

A realistic mass pattern is obtained if the bulk Majorana scale
is much smaller than the inverse of the compactification radius:
$|M|R \ll 1$.
In such a regime, the parameters $r$ and $|M|R$ are related as
\begin{eqnarray}
\frac{m_1^2-m_3^2}{m_2^2-m_1^2}
\,\simeq \,
\frac{\pm r}{4\pi |M|R(r^2-1)},
\label{Ratio}
\end{eqnarray}
where the plus (minus) sign in the right-hand side is for $r>1$~($r<1$).
By using~(\ref{Ratio}), the magnitude of $\theta_{12}$ is estimated as
\begin{eqnarray}
\tan2\theta_{12}
\,\simeq\,
4\frac{|\Delta m^2_{31}|}{\Delta m^2_{21}}\left| \frac{r^2-1}{r^2+1} \right|. 
\end{eqnarray}
The angle $\theta_{12}$ is zero for $r=1$ and it rapidly increases
as $r$ departs from unity due to the large factor 
$|\Delta m^2_{31}|/\Delta m^2_{21}\simeq 30$.
That is, the solar angle is small at the degenerate limit while 
it becomes large as the spectrum approaches the inverted hierarchy.

Figure~\ref{fig3} shows $\sin^2\theta_{12}\mathsf{}$ as a function of the total 
neutrino mass $\Sigma = \sum_i m_i$.
The blue (shaded) region is for the category ${\cal B}$ and
the pink (light-shaded) region is for ${\cal C}$.
The horizontal band is the 3$\sigma$ range of $\sin^2\theta_{12}$~\cite{3nu}.
The thickness of each plot corresponds to the possible 3$\sigma$ values of 
$|\Delta m^2_{31}|/\Delta m^2_{21}$ in the same reference.

In the case ${\cal B}$, the solar angle is increased as the spectrum is
shifted from the degenerate pattern to the hierarchical one.
The 3$\sigma$ range of $\sin^2 \theta_{12}=0.27 - 0.38$ is reproduced
for $0.10\,{\rm eV}\lesssim \Sigma \lesssim 0.12\,{\rm eV}$, which range will be tested 
in the cosmological observation such as cosmic microwave background data from the 
Planck mission.
While in the case ${\cal C}$, the solar angle is increased as the total mass
is increased.
This opposite behavior is because of the $45^\circ$ difference of the solar
angle $\theta_{12}$.
The predicted range of $\Sigma$ is slightly wider than the case of ${\cal B}$ 
and shifted to a higher region:
$0.11\,{\rm eV}\lesssim \Sigma \lesssim 0.15\,{\rm eV}$.

In this model, $\theta_{13}$ and $\theta_{23}$ comes from the charged-lepton 
sector and their magnitudes depend on the structure of the charged-lepton mass matrix.
The charged-lepton mass matrices suitable for the two options of the neutrino 
sector~(Model~I and I\!I ) are discussed in the next subsection.

\medskip

\subsection{The charged-lepton sector}\noindent
The charged-lepton sector resides on the SM boundary $y=\pi R$.
The right-handed electron $e_R$ and the pair of muon and tauon $(\mu_R,\tau_R)$ 
are assigned to be $\underline{1}$ and $\underline{2}$, respectively.
We also introduce new gauge singlet scalars 
$\phi_i$, which are assigned to be $\underline{3}$.
These assignments are summarized in the table below:
\vspace{3mm}
\begin{center}
\begin{tabular}{|c|ccc|c|c|}
\hline 
& $e_R$ & $(\mu_R,\tau_R)$ & $(L_e,L_\mu ,L_\tau )$ & $H$ & 
$(\phi _1,\phi _2,\phi _3)$ \\
\hline 
$S_4$ & $\underline{1}$ & $\underline{2}$ & $\underline{3}$ & $\underline{1}$ 
& $\underline{3}$ \\
\hline 
\end{tabular}
\end{center}
\vspace{3mm}
For the charged leptons, the $S_4$ invariant Lagrangian is 
\begin{align}
\mathcal{L}
&\,=\,\frac{Y_s}{\Lambda} \, \overline{e_R}\left 
(L_e\phi _1+L_\mu \phi _2+L_\tau \phi _3\right )H^* \nonumber \\
&\,\, + \frac{Y_d}{\Lambda} \left[\, \left(
\overline{\mu_R}+\overline{\tau_R}\right )L_e\phi_1
+\left(\omega^2\overline{\mu_R}+\omega \overline{\tau_R}\right )L_\mu \phi_2
+\left(\omega\overline{\mu_R}+\omega^2\overline{\tau_R}\right)L_\tau \phi _3
\, \right]H^*,
\end{align}
where $Y_s$, $Y_d$ are the dimensionless Yukawa couplings.
After the electroweak symmetry breaking,
the charged-lepton mass matrix is given by 
\begin{equation}
{M_\ell }=
v Y_s\begin{pmatrix}
\alpha _1 & \alpha _2 & \alpha _3 \\
0 & 0 & 0 \\
0 & 0 & 0
\end{pmatrix}+ v Y_d
\begin{pmatrix}
0 & 0 & 0 \\
\alpha _1 & \omega ^2\alpha _2 & \omega \alpha _3 \\
\alpha _1 & \omega \alpha _2 & \omega ^2\alpha _3
\end{pmatrix},
\end{equation}
where $\alpha_i \equiv \langle \phi _i\rangle /\Lambda $.
The Hermitian matrix $M_\ell^\dagger M_\ell $ becomes
\begin{equation}
M_\ell ^\dagger M_\ell =v^2
\begin{pmatrix}
\left (|Y_s|^2+2|Y_d|^2\right )\alpha _1^2 & 
\left(|Y_s|^2-|Y_d|^2\right )\alpha_1\alpha_2 & 
\left (|Y_s|^2-|Y_d|^2\right )\alpha_1\alpha_3 \\
\left (|Y_s|^2-|Y_d|^2\right )\alpha_1\alpha_2 & 
\left(|Y_s|^2+2|Y_d|^2\right )\alpha_2^2 & 
\left (|Y_s|^2-|Y_d|^2\right )\alpha_2\alpha_3 \\
\left (|Y_s|^2-|Y_d|^2\right )\alpha_1\alpha_3 & 
\left(|Y_s|^2-|Y_d|^2\right )\alpha_2
\alpha_3 & \left(|Y_s|^2+2|Y_d|^2\right )\alpha_3^2
\end{pmatrix}.
\label{mch1}
\end{equation}
By assuming that $\alpha_1 v\sim m_e$, $\alpha_2 v\sim m_\mu$, $\alpha_3 v\sim m_\tau$ 
and the Yukawa couplings are of order unity, one obtains the realistic charged-lepton 
masses and small mixing in~(\ref{mch1}).
This case is suitable for Model~I for the neutrino sector
because the near tri-bimaximal mixing comes from the neutrino sector.
Corrections from the charged-lepton mass matrix are at most ${\cal O}(m_e/m_\mu)$,
${\cal O}(m_e/m_\tau)$ and ${\cal O}(m_\mu/m_\tau)$
for the lepton mixing angles $\theta_{12}$, $\theta_{13}$ and $\theta_{23}$,
respectively.

On the other hand, the neutrino sector of Model~I\!I provides only $\theta_{12}$ 
and needs (at least) a large 2-3 mixing arising from the charged-lepton mass matrix.
This is possible if $\alpha _1=0$ and $\alpha _2=\alpha _3=\alpha $ are realized; 
\begin{equation}
M_\ell ^\dagger M_\ell =v^2
\begin{pmatrix}
0 & 0 & 0 \\
0 & \left (|Y_s|^2+2|Y_d|^2\right )\alpha^2 & \left(|Y_s|^2-|Y_d|^2\right )\alpha^2 \\
0 & \left (|Y_s|^2-|Y_d|^2\right )\alpha^2 & \left(|Y_s|^2+2|Y_d|^2\right )\alpha^2
\end{pmatrix}.
\end{equation}
The charged-lepton masses and the mixing angle are 
\begin{equation}
m_e^2=0,\quad  m_\mu ^2=3|Y_d|^2\alpha ^2v^2, \quad
m_\tau^2=\left (2|Y_s|^2+|Y_d|^2\right )\alpha ^2v^2,\quad
\theta _{23}^\ell= \frac{\pi}{4}.
\end{equation}
By taking $Y_d/Y_s\simeq \sqrt{2/3}\,m_\mu/m_\tau$, 
one obtains the observed mass ratio $m_\mu/m_\tau$ and the 
large 2-3 angle which accounts for the atmospheric neutrino oscillation.
For the Yukawa coupling $Y_s$ of order unity, $\alpha  =\mathcal{O}(10^{-2})$.
The nonzero electron mass is produced by holding a finite 
but small value of $\alpha_1$.

Another interesting way to obtain the realistic electron mass is to 
take account of the higher-order corrections to the charged-lepton mass matrix
while holding the vacuum-alignment $\alpha _1=0$ and $\alpha _2=\alpha _3=\alpha$ 
strictly.
The Lagrangian with the dimension six operators is given by
\begin{align}
\Delta \mathcal{L}
&\,=\,\frac{Y_s^\Delta}{\Lambda^2} \overline{e_R}\left(\, L_e\phi_2\phi_3
\,+\, L_\mu \phi_1\phi_3 \,+\, L_\tau \phi_1\phi_2 \, \right)H^* \nonumber \\
&\,\, + \frac{Y_d^\Delta}{\Lambda^2} 
\left[\, \left(\overline{\mu_R}+\overline{\tau_R}\right)L_e\phi_2\phi_3
\,+\, \left(\omega^2\overline{\mu_R}+ \omega\overline{\tau_R}\right)L_\mu \phi_1\phi_3
\,+\, \left(\omega\overline{\mu_R}+\omega^2\overline{\tau_R}\right )L_\tau\phi_1\phi_2
\, \right] H^*.
\end{align}
After $\phi_i$ developing the VEVs of 
$\langle \phi_i \rangle = (0,\alpha\Lambda,\alpha\Lambda)$,
the charged-lepton mass matrix up to 
this order is written as 
\begin{equation}
\hat M_\ell =M_\ell +\Delta M_\ell =
v \begin{pmatrix}
Y_s^\Delta \alpha^2 & Y_s \alpha & Y_s\alpha \\
Y_d^\Delta \alpha^2 & Y_d \omega ^2\alpha & Y_d \omega \alpha \\
Y_d^\Delta \alpha^2 & Y_d \omega \alpha & Y_d \omega^2\alpha 
\end{pmatrix}.
\end{equation}
The Hermitian matrix $\hat M_\ell ^\dagger \hat M_\ell $ follows
\begin{equation}
\hat M_\ell ^\dagger \hat M_\ell =\alpha ^2v^2
\begin{pmatrix}
\left(|Y_s^\Delta|^2+2|Y_d^\Delta|^2\right )\alpha ^2 & 
\left(Y_s^\Delta{}^*Y_s - Y_d^\Delta{}^* Y_d \right )\alpha & 
\left(Y_s^\Delta{}^*Y_s - Y_d^\Delta{}^* Y_d \right )\alpha \\[.5mm]
\left(Y_s^\Delta Y_s^* - Y_d^\Delta Y_d^* \right )\alpha &  
 |Y_s|^2+2|Y_d|^2 & |Y_s|^2-|Y_d|^2 \\[.5mm]
\left(Y_s^\Delta Y_s^* - Y_d^\Delta Y_d^* \right )\alpha  
& |Y_s|^2-|Y_d|^2 & |Y_s|^2+2|Y_d|^2\\
\end{pmatrix}.
\end{equation}
The charged-lepton masses and mixing are given by
\begin{equation}
m_e^2 \,\simeq\, 2|Y_d^\Delta|^2 \alpha ^4v^2, \qquad 
m_\mu ^2\, \simeq\,  3|Y_d|^2\alpha ^2v^2, \qquad 
m_\tau^2\,\simeq\, \left(2|Y_s|^2+|Y_d|^2\right)\alpha ^2v^2, 
\end{equation}
\begin{equation}
\theta _{12}^\ell=0,  \qquad
\theta _{13}^\ell \simeq \frac{\sqrt{2}\left| Y_s^\Delta Y_s^*-Y_d^\Delta Y_d^*\right|
\alpha }{2|Y_s|^2+|Y_d|^2},
\qquad 
\theta _{23}^\ell= \frac{\pi}{4},
\end{equation}
where the 1-3 angle $\theta_{13}^\ell$ is assumed to be small not to 
contradict the reactor bound.
As in the case of the leading-order estimation, 
$Y_d/Y_s\simeq \sqrt{2/3}\,m_\mu/m_\tau$ and 
$\alpha  =\mathcal{O}(10^{-2})$ for $Y_s \simeq 1$. 
The electron mass is in the right range if $Y_d^\Delta \approx Y_d$.
The 1-3 angle becomes $\theta _{13}^\ell \simeq 1/\sqrt{2} (Y_s^\Delta /Y_s)$
which will be observable if $Y_s^\Delta /Y_s = \mathcal{O}(10^{-1})$.

A possible way to achieve the aligned vacuum $\langle \phi_i \rangle = 
(0,\alpha\Lambda,\alpha\Lambda)$ is to assume that the scalar fields 
$\phi_i$ propagate in the five-dimensional bulk and follow nontrivial boundary
conditions~\cite{TFS}.
By assigning the pseudotriplet $\underline{3}'$ to $\phi_i$ or
assuming the scalar potential to have additional $Z_2$ symmetry, it is possible to 
impose the twisted boundary condition $\phi(y + 2\pi R) = -d_2 \phi(y)$.
While the first component $\phi_1$ cannot develop a constant VEV because of the 
antiperiodic boundary condition, the other two components $\phi_2$ and $\phi_3$ 
acquire equal VEVs since the boundary condition respects $S_2$ symmetry which is 
the exchange of the 2-3 component of the triplet $\phi_i$.

The bulk scalars $\phi_i$ with nontrivial boundary conditions are also useful
for realizing a hierarchal charged-lepton mass matrix suitable for the Model~I.
Suppose that the right-handed charged-leptons $(e_R, \mu_R,\tau_R)$ are 
$\underline{3}$ of $S_4$ and also propagate in the bulk.

The gauge singlet scalars $\phi_i$ are also coupled to the neutrinos via 
higher-dimensional operators, e.g., $\overline{\Psi}LH\phi$. 
For the Model~I, the most stringent contribution comes from $\langle \phi_3 \rangle$ 
component related to the tau mass: $m_\tau \simeq \sqrt{|Y_s|^2 +2|Y_d|^2}v \alpha_3$.
However, $\alpha_3 =\mathcal{O}(10^{-2})$ if all the dimensionless couplings are 
of order unity.
Thus the corrections from such higher-dimensional operators
remain a few percent and the predictions discussed 
in Section~\ref{moA} and~\ref{moB} are valid unless the Yukawa couplings
are anomalously small.
A similar discussion is also hold for the Model~I\!I.

\section{Summary}
In this paper, we have examined the $S_4$ discrete group as a feasible
candidate for the family symmetry which accounts for the leptonic generation
structure.
Phenomenologically viable $S_4$ breaking is triggered by the boundary conditions 
of the right-handed neutrinos which reside on the five-dimensional spacetime.
With the SM fields being trapped in the four-dimensional subspace,
many boundary conditions produce realistic lepton mixing
slightly deviated from the tri-bimaximal mixing in a controllable way. 
There are two types of plausible structures (Model~I and~I\!I  in 
Section~\ref{moA} and~\ref{moB} respectively)
in the neutrino sector.

In the Model~I, the whole mixing matrix originates in the neutrino sector
and the charged-lepton mass matrix is diagonal in a certain $S_4$ basis (see 
Appendix~\ref{s4g}).
While the solar component $U_{e2}$ is robustly predicted to be $1/\sqrt{3}$,
the reactor angle $\theta_{13}$ is correlated to
the mass patterns.
With the normal mass ordering, $\theta_{13}$ increases as the total 
 neutrino mass decreases in such a way that $ \sin\theta_{13} \gtrsim 0.07$ 
if $m_1 + m_2 + m_3 \lesssim 0.06 \,{\rm eV}$, which range will be probed at
future experiments.
Furthermore, the atmospheric angle $\theta_{23}$ is deviated from the
maximal value $45^\circ$ in such a way that the deviation is ruled by the magnitude
of $\theta_{13}$.
While the discovery of $\theta_{13}$ is imminent for the normal mass hierarchy, 
the reactor angle is generally small ($\theta_{13} \lesssim 10^{-3}$) for the 
inverted mass ordering;  
the mixing matrix is predicted to be almost tri-bimaximal form in the case of
the inverted ordering.

While in the Model~I\!I, the large solar angle $\theta_{12}$ stems from the neutrino
sector and the maximal atmospheric angle $\theta_{23}$ emerges from the
charged-lepton mass matrix.
Unlike the Model~I, the Model~I\!I accommodates only the inverted mass ordering.
The solar angle $\theta_{12}$ is related to the total neutrino mass.
Within $3\sigma$ range of the current oscillation parameters, the total mass
is predicted to be in a small window $0.10 \,{\rm eV} \leq m_1 + m_2 + m_3 \leq 0.15
\,{\rm eV}$ which can be probed by the Planck satellite.
By the combined data of future oscillation experiments and cosmological observations,
the models discussed here are not only tested but also distinguished each other 
in their predictions for the leptonic generation structure.

Besides the predictions in the masses and the mixing angles discussed in this paper,
there are other phenomenological issues such as collider signals~\cite{collider}, 
leptogenesis~\cite{leptogen}, low-energy CP violation, dark matter, and so on.
In particular, it might be interesting to identify some of the KK right-handed neutrinos
as the dark matter by introducing the bulk Dirac masses with an appropriate
choice of the parameters.
These subjects remain to be explored in future work.

\medskip

\subsection*{Acknowledgments}
The authors are supported in part by the scientific grant from the
ministry of education, science, sports, and culture of Japan
(No.~21340055). 
H.I. is supported by Grand-in-Aid for Scientific Research, No.21.5817
from the Japan Society of Promotion of Science.
Y.S. is supported by Grand-in-Aid for Scientific Research, No.22.3014
from the Japan Society of Promotion of Science.
\bigskip

\appendix

\section{Lorentz spinors and gamma matrices}
\label{LG}
In this work, the gamma matrices are taken as
\begin{gather}
\{ \Gamma^M, \Gamma^N \} \,=\, 
2\eta^{MN} \,=\, 2\,{\rm diag}(+1, -1,-1,-1,-1), \\[2mm]
\Gamma_\mu = \gamma_\mu = 
\begin{pmatrix}
 & \sigma_\mu \\
 \bar{\sigma}_\mu & \\
\end{pmatrix},\qquad
i\Gamma_4 = \gamma_5 =  
\begin{pmatrix}
1 &  \\
  & -1 \\
\end{pmatrix},
\end{gather}
where $\sigma_\mu=(1,\sigma_i)$ and $\bar{\sigma}_\mu=(1,-\sigma_i)$.
A 4-component spinor is written in terms of 2-component spinors as
\begin{eqnarray}
\Psi \,=\, \begin{pmatrix}
\xi_\alpha \\
\eta^{\dot{\alpha}} \\
\end{pmatrix}.
\end{eqnarray}
The Dirac and charge conjugates for $\Psi$ are given by 
\begin{eqnarray}
\overline{\Psi} \,=\, 
\big(\eta^{*\alpha} \;\, \xi_{\dot{\alpha}}^* \big), \qquad\quad
\Psi^c \,=\, C_5 \overline{\Psi}^{\rm T} \,=\, \begin{pmatrix}
-\epsilon_{\alpha\beta}\eta^{*\beta} \\[.5mm]
-\epsilon^{\dot{\alpha}\dot{\beta}}\xi^*_{\dot{\beta}}\,
\end{pmatrix},
\end{eqnarray}
where $C_5$ is the charge conjugation matrix in five 
dimensions: $C_5=i\gamma^2\gamma^0\gamma_5$.
The antisymmetric tensors are
\begin{eqnarray}
\epsilon^{\alpha\beta} = \epsilon_{\alpha\beta} = 
\epsilon^{\dot{\alpha}\dot{\beta}} = \epsilon_{\dot{\alpha}\dot{\beta}}
 = \begin{pmatrix} & 1 \\ -1 &  \\ \end{pmatrix}.
\end{eqnarray}

\medskip

\section{$S_4$ group}
\label{s4g}
The $S_4$ group consists of all permutations of four objects and the order is $4! = 24$.
The $S_4$ is symmetry of the cube. 
The irreducible representations are two singlets $\underline{1}$ and $\underline{1}'$, 
one doublet $\underline{2}$, and two triplets $\underline{3}$ and $\underline{3}'$.
In this work, we adopt the following basis~\cite{dis}
\begin{eqnarray}
Q = \begin{pmatrix}0&1\\1&0\\\end{pmatrix},\quad
P = \begin{pmatrix}\omega&0\\0&\omega^2\\\end{pmatrix}
\quad\quad (\omega \equiv  e^{i\frac{2\pi}{3}})
\label{2}
\end{eqnarray}
for the doublet $\underline{2}$ and 
\begin{eqnarray}
Q = \begin{pmatrix}-1&0&0\\0&0&-1\\0&1&0\\\end{pmatrix},\quad
P = \begin{pmatrix}0&0&1\\1&0&0\\0&1&0\\\end{pmatrix}
\label{3}
\end{eqnarray}
for the triplet $\underline{3}$.
All the group elements are given by the products of these two generators;
\begin{eqnarray}
\begin{array}{llll}
a_1 = Q^4, &\quad a_2 = Q^2, &\quad a_3 = PQ^2P^2,&\quad a_4 = Q^2PQ^2P^2,\\
b_1 = P,&\quad b_2 = Q^2P, &\quad b_3 = QPQP^2, &\quad b_4 = Q^2PQ^2,\\
c_1 = P^2,&\quad c_2 = Q^2P^2,&\quad c_3 = QPQ, &\quad c_4 = Q^3PQ,\\
d_1 = PQPQ^2,&\quad d_2 = PQP, &\quad d_3 = Q^3, &\quad d_4 = Q,\\
e_1 = Q^2PQ,&\quad e_2 = PQ, &\quad e_3 = Q^3P^2, &\quad e_4 = QP^2,\\
f_1 = QPQ^2,&\quad f_2 = PQP^2, &\quad f_3 = P^2Q, &\quad f_4 = QP.\\
\end{array}
\end{eqnarray}
The tensor products $\underline{3} \times \underline{3}$ and 
$\underline{2} \times \underline{3}$ play an essential role to construct
the Yukawa coupling matrices in symmetric phase.
They are decomposed as $\underline{3} \times \underline{3} = \underline{1}
+ \underline{2} + \underline{3} + \underline{3}'$ and
$\underline{2} \times \underline{3} = \underline{3} + \underline{3}'$.
Suppose that $\psi_i$, $\phi_i$ $(i=1,2,3)$ and $\chi_j \,(j = 1,2)$ 
behave as $\underline{3}$ and $\underline{2}$ under the basis of~(\ref{3}) 
and~(\ref{2}), respectively.
Then it follows
\begin{eqnarray}
\psi \times \phi &=& 
\underbrace{\sum_{i =1}^3 \psi_i\phi_i}_{\underline{1}} 
+ 
\underbrace{
\left( \begin{array}{c}
\psi_1 \phi_1 + \omega \psi_2 \phi_2 + \omega^2 \psi_3 \phi_3 \\
\psi_1 \phi_1 + \omega^2 \psi_2 \phi_2 + \omega \psi_3 \phi_3 \\
\end{array}
\right)
}_{\underline{2}}
\nonumber\\
&&\quad\quad\quad\quad+
\underbrace{
\left( \begin{array}{c}
\psi_2 \phi_3 + \psi_3 \phi_2 \\
\psi_3 \phi_1 + \psi_1 \phi_3\\
\psi_1 \phi_2 + \psi_2 \phi_1\\
\end{array}
\right)
}_{\underline{3}}
+
\underbrace{
\left( \begin{array}{c}
\psi_2 \phi_3 - \psi_3 \phi_2 \\
\psi_3 \phi_1 - \psi_1 \phi_3\\
\psi_1 \phi_2 - \psi_2 \phi_1\\
\end{array}
\right),
}_{\underline{3}'}\\
\chi \times \psi &=& 
\underbrace{
\left( \begin{array}{c}
(\chi_1 + \chi_2) \psi_1 \\
(\omega^2 \chi_1 + \omega \chi_2) \psi_2 \\
(\omega \chi_1 + \omega^2 \chi_2) \psi_3 \\
\end{array}
\right)
}_{\underline{3}}
+
\underbrace{
\left( \begin{array}{c}
(\chi_1 - \chi_2) \psi_1 \\
(\omega^2 \chi_1 - \omega \chi_2) \psi_2 \\
(\omega \chi_1 - \omega^2 \chi_2) \psi_3 \\
\end{array}
\right).
}_{\underline{3}'}
\end{eqnarray}


\end{document}